\documentclass{ws-mpla-notrim}

\def\lQ{\Lambda_{\rm QCD}}

\newcommand{\nn}{\nonumber}
\newcommand{\be}{\begin{equation}}
\newcommand{\ee}{\end{equation}}
\newcommand{\bea}{\begin{eqnarray}}
\newcommand{\eea}{\end{eqnarray}}
\def\al{\alpha}
\def\als{\alpha_{\rm s}}
\def\siml{{\ \lower-1.2pt\vbox{\hbox{\rlap{$<$}\lower6pt\vbox{\hbox{$\sim$}}}}\ }} 
\def\simg{{\ \lower-1.2pt\vbox{\hbox{\rlap{$>$}\lower6pt\vbox{\hbox{$\sim$}}}}\ }} 
\newcommand{\MS}{\overline{\rm MS}}

\begin{document}

\markboth{Antonio Vairo}{A theoretical review of heavy quarkonium inclusive decays}

%
\catchline{}{}{}{}{}
%

\title{A THEORETICAL REVIEW OF HEAVY QUARKONIUM INCLUSIVE DECAYS}

\author{\footnotesize ANTONIO VAIRO}

\address{Dipartimento di Fisica dell'Universit\`a di Milano, 
  via Celoria 16, 20133 Milan, Italy \\
     antonio.vairo@mi.infn.it}

\maketitle


\begin{abstract}
In this brief review, I summarize the current theoretical knowledge of 
heavy quarkonium inclusive decays, with emphasis on recent progress 
made in the framework of QCD effective field theories.
In appendix, I list the imaginary parts of the matching coefficients  
of the dimension 6 and dimension 8 NRQCD four-fermion operators as presently known.
\end{abstract}

\ccode{PACS Nos.: 12.38.-t, 12.39.St, 13.25.Gv}

\section{Introduction}
Heavy quarkonia (charmonium, bottomonium, ...) provide an ideal set of 
observables to probe properties of low-energy QCD in a controlled way. 
The reason is the following. Heavy quarkonia are non-relativistic 
bound states and, therefore, characterized by a set of energy scales
hierarchically ordered: $m$, $mv$, $mv^2$, ... where  $m$ is the heavy-quark 
mass and $v\ll 1$ the relative heavy-quark velocity. For heavy quarkonia, 
$m$ is much larger than the scale of non-perturbative physics, $\lQ$,  
and, therefore, degrees of freedom associated with that scale can be treated 
perturbatively and calculations done order by order in $\als$. 
The non-relativistic hierarchy of scales also survives below $\lQ$.
Therefore, for any heavy quarkonium state the low-energy dynamics is organized 
in matrix elements ordered in powers of $v$ (and, in general, $\lQ/m$). To any given order in $\als$ and $v$, 
only a finite number of Feynman diagrams and matrix elements respectively have to be calculated.

The way to implement rigorously these expansions in QCD is provided 
by the non-relativistic effective field theories (EFTs) of QCD. 
The first has been Non-Relativistic QCD, NRQCD\cite{nrqcd0,nrqcd}. It is obtained 
from QCD by integrating out degrees of freedom of energy $m$.
NRQCD still contains the lower energy scales as dynamical degrees of freedom. 
In the last few years, the problem of integrating out the remaining dynamical scales has been addressed 
by several groups and has now reached a solid level of understanding 
(lists of references may be found in\cite{reveft}). 
The ultimate EFT obtained by subsequent matchings from QCD, where only the
lightest degrees of freedom of energy $mv^2$ are left dynamical, 
is potential NRQCD, pNRQCD\cite{Mont,long}. This EFT is close to a
quantum-mechanical description of the bound system and, therefore, as simple. 
It has been systematically explored in the dynamical
regime $mv^2 \simg \lQ$ in\cite{long,logs} and in the regime $mv^2 \ll \lQ$ 
in\cite{long,M12,pw,sw,sqrt}. An alternative approach to pNRQCD has been suggested 
in\cite{vnrqcd}. This approach has been developed in the dynamical situation 
$mv^2 \gg \lQ$, but not been extended, so far, to $mv^2 \gg\!\!\!\!\!\!/ \; \lQ$ 
where most of the heavy quarkonium states are believed to lie. 

In this letter, I will review the theory status of heavy quarkonium inclusive 
and electromagnetic decays into light particles in the framework of QCD non-relativistic EFTs.
The main mechanism of heavy quarkonium decay into light particles is quark--anti-quark
annihilation. Since this happens at a scale $2m$, which is perturbative, the heavy quarks 
annihilate into the minimal number of gluons allowed by symmetry. 
Experimentally this fact is reflected by the narrow width of the 
heavy quarkonia below the open flavour threshold. 
In an EFT language, once the scale $m$ has been integrated out, the information 
on decays is carried by contact terms (four-fermion operators) whose matching coefficients 
develop an imaginary part. The low-energy dynamics is in the matrix elements 
of the four-fermion operators evaluated on the heavy quarkonium states.
If one assumes that only heavy quarkonium states with quark--anti-quark in a singlet 
configuration can exist, then only singlet four-fermion operators contribute and 
the matrix elements reduce to heavy quarkonium wave functions (or derivatives of them) 
calculated in the origin. This assumption is known as the ``singlet model''.
Explicit calculations show that at higher order the singlet matching 
coefficients develop infrared divergences (for $P$ waves 
this happens at next-to-leading order\cite{IR}: compare with the expressions in \ref{appA}). 
In the singlet model, these do not cancel 
in the expression of the decay widths. It has been the first success 
of NRQCD to show that, due to the non-Abelian nature of QCD, the Fock space of a heavy quarkonium 
state may contain a small component of quark--anti-quark in an octet configuration bound 
with some gluonic degrees of freedom (the component is small because operators coupling
transverse gluons with quarks are suppressed by powers of $v$), 
due to this component, matrix elements of octet four-fermion operators contribute 
and, finally, exactly these contributions absorbe the infrared divergences 
of the singlet matching coefficients in the decay widths, giving rise to finite results\cite{BBL0,nrqcd}.
NRQCD is now the standard framework to study heavy quarkonium decays.
From the theoretical side in recent years the main effort has gone into two obvious directions: 
(1) improving the knowledge of the perturbative 
series of the matching coefficients either by fixed order calculations 
or by resumming large contributions (renormalons or large logs); 
(2) improving the knowledge of the NRQCD matrix elements either by direct 
evaluation, which may be obtained by fitting the experimental data, by lattice 
calculations, and by models, or by exploiting the hierarchy of scales still entangled 
in NRQCD and constructing EFTs of lower energy. I already mentioned 
that pNRQCD is the ultimate of these EFTs. In such context a 
new factorization can be achieved that allows to reduce, under 
some dynamical circumstances, the number of non-perturbative parameters 
in the expression of the decay widths.

Experimentally several facilities have operated and are operating 
in an energy range relevant for heavy quarkonium. I refer to the pages 
of the Quarkonium Working Group for a broad overview\cite{qwg}. 
Here, I would like to mention only some of the data produced in the last 
few years relevant for heavy quarkonium inclusive and electromagnetic decays.
They come from the E835 experiment at FNAL, where heavy quarkonium 
is produced from $p\,\bar{p}$ annihilation (E835 operated in the charmonium energy region)  
and from the B-factories (BABAR at SLAC and BELLE at KEKB, which 
operate at the $\Upsilon(4S)$ resonance), BES at BEPC (operating in the charmonium 
energy region) and CLEO at CESR (CLEOIII took data in the bottomonium energy region, 
CLEO-c is taking data in the charmonium one) where heavy quarkonium is produced from  $e^+e^-$ collisions.
In the bottomonium system, CLEO has provided the first experimental values (still affected 
by large uncertainties) for the ratios of the inclusive decay widths 
of $2P_J$ bottomonium states, extracted from the data on 
two-photons transitions from $\Upsilon(3S)$ decays\cite{CLEOchib}. 
In the charmonium system, new determinations of the $\eta_c$ resonance parameters 
came from the experiments E835\cite{E8352}, BABAR\cite{BaBar} (where the 
$\eta_c$ is produced from two-photon interactions), BELLE\cite{Belleetac,Belle2} 
($B\to\eta_c \, K$) and BES\cite{BES2} ($J/\psi \to \eta_c\,\gamma$).
I refer to\cite{skw} for comparison and discussion of the data. 
E835 also provides $\Gamma(\eta_c \to \gamma \, \gamma)$\cite{E8352}. 
The resonance parameters of the $\psi(2S)$ have been newly extracted from high 
statistics $e^+e^-$ cross-section data at BES\cite{BES}. 
The same collaboration also reports a new determination of 
$\Gamma(\psi(2S)\to e^+e^-)$. The branching ratio for $\psi(2S)\to e^+e^-$ has been 
measured by the E835 experiment\cite{E835psi2s}. The branching ratios  of 
$\psi(2S)\to e^+e^-$ and $\psi(2S)\to \mu^+\mu^-$ have been measured by the BABAR 
experiment\cite{BaBarpsi2s}. BELLE reported the first observation of the $\eta_c(2S)$  
and the measurement of its resonance parameters (from $B \to \eta_c(2S) \, K$
in\cite{Belle2} and $e^+e^- \to J/\psi\, \eta_c(2S)$ in\cite{Belle3}) 
followed by BABAR\cite{BaBar} and CLEO\cite{CLEOetac2} 
(in both cases from two-photon interactions).
For what concerns the $L=1$ charmonium states, the $\chi_{c0}$ resonance 
parameters have been newly measured at E835\cite{E835}.
The same collaboration provides a determination of 
$\Gamma(\chi_{c0}\to \gamma\,\gamma)$\cite{graphd,mussa}
and $\Gamma(\chi_{c2}\to \gamma\,\gamma)$\cite{E835chi2}.
New values for $\Gamma(\chi_{c0}\to \gamma\,\gamma)$ and 
$\Gamma(\chi_{c2}\to \gamma\,\gamma)$ have been provided by CLEO\cite{CLEO1}
and for $\Gamma(\chi_{c2}\to \gamma\,\gamma)$ by BELLE\cite{Belle} in both cases 
from two-photon production processes.

The letter is organized as follows. In Sec. \ref{secnrqcd}, I review the NRQCD factorization 
formulas for heavy quarkonium decay widths. In Sec. \ref{secper}, I briefly discuss the 
perturbative series of the matching coefficients. \ref{appA} 
contains a complete list of all the imaginary parts of the matching coefficients 
of the dimension 6 and 8 operators (hadronic and electromagnetic)  at their 
present accuracy. In Sec. \ref{secrel}, I discuss the NRQCD matrix elements 
and in Sec. \ref{secpnrqcd} the pNRQCD factorization.
Some final remarks are given in Sec. \ref{seccon}.

\section{NRQCD}
\label{secnrqcd}
The NRQCD factorization formulas are obtained by separating contributions coming from 
degrees of freedom of energy $m$ from those coming from  degrees of freedom of lower 
energy. In the case of heavy quarkonium decay widths, they have been 
rigorously proved\cite{nrqcd}.\footnote{Such a proof is still lacking for 
the NRQCD factorization of heavy quarkonium production cross sections.}
High-energy contributions are encoded into the imaginary parts of the four-fermion
matching coefficients, $f,g_{1,8,ee,\gamma\gamma}(^{2S+1}L_J)$ and are ordered 
in powers of $\als$. Low-energy contributions are encoded into the matrix elements 
of the four-fermion operators on the heavy quarkonium states $|H\rangle$ 
($\langle \dots \rangle_{H} \equiv \langle H  | \dots  | H \rangle$). 
These are, in general, non-perturbative objects, which can scale as
powers of $\lQ$, $mv$, $mv^2$, ... (i.e. of the low-energy dynamical scales of NRQCD). 
Therefore, matrix elements of higher dimensionality are suppressed by powers of $v$ or $\lQ/m$.
Including up to the NRQCD four-fermion operators of dimension 8, the NRQCD factorization formulas 
for inclusive decay widths of heavy quarkonia into light hadrons ($LH$) read\cite{nrqcd,BBL0}:
\bea
&&\Gamma(V_Q (nS) \rightarrow LH) = {2\over m^2}\Bigg( 
{\rm Im\,}f_1(^3 S_1) \,  \langle O_1(^3S_1)\rangle_{V_Q(nS)}
\nn
\\
&&\quad
+ {\rm Im\,}f_8(^3 S_1)\, \langle O_8(^3S_1)\rangle_{V_Q(nS)}
+ {\rm Im\,}f_8(^1 S_0)\, \langle O_8(^1S_0)\rangle_{V_Q(nS)} 
\nn
\\
&&\quad 
+ {\rm Im\,}g_1(^3 S_1)\,
{\langle {\mathcal P}_1(^3S_1)\rangle_{V_Q(nS)} \over m^2}
+ {\rm Im\,}f_8(^3 P_0)\,
{\langle O_8(^3P_0)\rangle_{V_Q(nS)} \over m^2}
\nn
\\
&&\quad
+ {\rm Im\,}f_8(^3 P_1)\,
{\langle O_8(^3P_1)\rangle_{V_Q(nS)} \over m^2}
+ {\rm Im\,}f_8(^3 P_2)\,
{\langle O_8(^3P_2)\rangle_{V_Q(nS)} \over m^2}\Bigg),
\label{gamma1}
\\
&&
\nn 
\\
&&\Gamma(P_Q (nS) \rightarrow LH) = {2\over m^2}\Bigg( 
{\rm Im\,}f_1(^1 S_0)\,   \langle O_1(^1S_0)\rangle_{P_Q(nS)}
\nn
\\
&&\quad
+ {\rm Im\,}f_8(^1 S_0)\, \langle O_8(^1S_0)\rangle_{P_Q(nS)}
+ {\rm Im\,}f_8(^3 S_1)\, \langle O_8(^3S_1)\rangle_{P_Q(nS)} 
\nn
\\
&&\quad 
+ {\rm Im\,}g_1(^1 S_0)\,
{\langle {\mathcal P}_1(^1S_0)\rangle_{P_Q(nS)} \over m^2}
+ {\rm Im\,}f_8(^1 P_1)\,
{\langle O_8(^1P_1)\rangle_{P_Q(nS)} \over m^2} \Bigg),
\\
&&
\nn
\\
&&\Gamma(\chi_Q(nJS)  \rightarrow LH)= 
{2\over m^2}\Bigg( {\rm Im \,}  f_1(^{2S+1}P_J)\, 
{\langle O_1(^{2S+1}P_J ) \rangle_{\chi_Q(nJS)} \over m^2}
\nn
\\
&&\quad
+  {\rm Im \,} f_8(^{2S+1}S_S) \,\langle O_8(^1S_0 ) \rangle_{\chi_Q(nJS)} \Bigg).
\eea
At the same order the electromagnetic decay widths are given by:
\bea 
&&\Gamma(V_Q (nS) \rightarrow e^+e^-)= {2\over m^2}\Bigg( 
{\rm Im\,}f_{ee}(^3 S_1)\,   \langle O_{\rm EM}(^3S_1)\rangle_{V_Q(nS)}
\nn
\\
&&\quad
+ {\rm Im\,}g_{ee}(^3 S_1)\,
{\langle {\mathcal P}_{\rm EM}(^3S_1)\rangle_{V_Q(nS)} \over m^2}\Bigg),
\\
&&
\nn
\\
&&\Gamma(P_Q (nS) \rightarrow \gamma\gamma)= {2\over m^2}\Bigg( 
{\rm Im\,}f_{\gamma\gamma}(^1 S_0)\,   
\langle O_{\rm EM}(^1S_0)\rangle_{P_Q(nS)}
\nn
\\
&&\quad
+ {\rm Im\,}g_{\gamma\gamma}(^1 S_0)\,
{\langle {\mathcal P}_{\rm EM}(^1S_0)\rangle_{P_Q(nS)} \over m^2} \Bigg),
\\
&&
\nn
\\
&&\Gamma(\chi_Q(nJ1)  \rightarrow \gamma\gamma)= 
2 \, {\rm Im \,}  f_{\gamma\gamma}(^3P_J)\, 
{\langle O_{\rm EM}(^3P_J )\rangle_{\chi_Q(nJ1)}  \over m^4}, 
\quad J=0,2\,.
\label{chigg}
\label{gammachi}
\eea
The symbols $V_Q$ and $P_Q$ indicate respectively the vector and pseudoscalar $S$-wave heavy 
quarkonium and the symbol $\chi_Q$ the generic $P$-wave quarkonium (the states
$\chi_Q(n10)$ and $\chi_Q(nJ1)$ are usually called $h_Q((n-1)P)$ and
$\chi_{QJ}((n-1)P)$, respectively). 

The operators $O,{\mathcal P}_{1,8,{\rm EM}}(^{2S+1}L_J)$ are the dimension $6$ and  $8$ 
four-fermion operators of the NRQCD Lagrangian. They are classified in dependence of their 
transformation properties under colour as singlets ($1$) and octets ($8$) and
under spin ($S$), orbital ($L$) and total angular momentum ($J$). 
The operators with the subscript EM are the singlet operators projected on 
the QCD vacuum. The explicit expressions of the operators can be found in\cite{nrqcd}
(or listed in \ref{appA} of\cite{sw}).

\section{The perturbative expansion}
\label{secper}
The imaginary parts of the four-fermion matching coefficients have been
calculated over the past twenty years by different authors and 
to different levels of precision. Since the results are scattered over a 
large number of papers, some of them being difficult to collect,  
some having been corrected in subsequent publications
and some still being in disagreement with each other, I have 
listed all the imaginary parts of the matching coefficients of the dimension 6 and 8 operators 
(hadronic and electromagnetic) at the present accuracy in \ref{appA}.
The tree-level matching of the dimension 10 $S$-wave operators can be found in\cite{bope}. 
The tree-level matching of the dimension 9 electromagnetic $P$-wave operators can be found in\cite{mawang}.

The convergence of the perturbative series of the four-fermion matching coefficients 
is often bad. Let us consider, for instance, the following matching coefficients 
($n_f = 3$, $\mu_R = 2m$)\cite{conf5}:
\begin{eqnarray*}
&&
{\rm Im}f_{1}(^1S_0)  = (\dots)\times \left(1 + 11.1 \, {\als 
  \over \pi}\right), \\
&&
{\rm Im}f_{8}(^1S_0)  = (\dots)\times \left(1 + 13.7 \, {\als 
  \over \pi}\right), \\
&&
{\rm Im}f_{8}(^3S_1)  = (\dots)\times \left(1 + 10.3 \, {\als 
  \over \pi}\right), \\
&&
{\rm Im}f_{1}(^3P_0)  = (\dots)\times \left(1 + \left(13.6 - 0.44 \, \log {\mu\over 2m} \right)
\, {\als \over \pi}\right), \\
&&
{\rm Im}f_{1}(^3P_2)  = (\dots)\times \left(1 - \left(0.73  + 1.67 \, \log {\mu\over 2m} \right) 
\, {\als \over \pi}\right). 
\end{eqnarray*}
Apart from the case of ${\rm Im}f_{1}(^3P_2)$, the series in $\als$ of the other
coefficients does not show convergence.
This behaviour cannot be adjusted by a suitable choice of the factorization scale $\mu$, which enters
only in ${\rm Im}f_{1}(^3P_{0,2})$.
A solution may be provided by the resummation of the large contributions in the perturbative series 
coming from bubble-chain diagrams. This analysis has been successfully carried
out for $S$-wave annihilation decays\cite{chen}.
A treatment that includes $P$-wave decays is still missing.

\section{The relativistic expansion}
\label{secrel}
The NRQCD matrix elements may be fitted on the experimental decay data\cite{Maltoni,mussa} or
calculated on the lattice\cite{latbo}. The matrix elements of singlet
operators can be linked at leading order to the Schr\"odinger wave functions in the
origin\cite{nrqcd} and therefore may be evaluated by means of potential models\cite{EichtenQuigg}.
In general, however, NRQCD matrix elements, in particular of higher dimensionality, 
are poorly known or completely unknown. 

It has been discussed in\cite{mawang} and\cite{bope}, 
that higher-order operators, not included in the 
formulas of Sec. \ref{secnrqcd}, even if parametrically suppressed,  
may turn out to give sizeable contributions to the decay widths. 
This may be the case, in  particular, 
for charmonium, where $v^2 \sim 0.3$, so that relativistic corrections 
are large, and for $P$-wave decays where the above formulas provide,
indeed, only the leading-order contribution in the velocity expansion.
In fact it was pointed out in\cite{mawang,conf5} that if no special cancellations  
among the matrix elements occur, then the order $v^2$ relativistic corrections 
to the electromagnetic decays $\chi_{c0} \to \gamma\gamma$ and $\chi_{c2}\to \gamma\gamma$ 
may be as large as the leading terms.

In\cite{Maltoni} it was also noted that the numerical relevance of higher-order 
matrix elements may be enhanced by the multiplying matching coefficients.
This is, indeed, the case for the decay width of $S$-wave vector states, 
where the matching coefficients multiplying the octet matrix elements (with the only exception 
of ${\rm Im} f_8(^3P_1)$) are enhanced by $\als$ with respect 
to the coefficient ${\rm Im} f_1(^3S_1)$ of the leading singlet matrix element  
(see Eq. (\ref{gamma1}) and \ref{appA}).

In the bottomonium system, 14 $S$- and $P$-wave states lie below the open flavour threshold 
($\Upsilon(nS)$ and $\eta_b(nS)$ with $n=1,2,3$; $h_b(nP)$ and $\chi_{bJ}(nP)$ 
with $n=1,2$ and $J=0,1,2$) and in the charmonium system 8 
($\psi(nS)$ and $\eta_c(nS)$ with $n=1,2$; $h_c(1P)$ and
$\chi_{cJ}(1P)$ with $J=0,1,2$). For these states 
Eqs. (\ref{gamma1})-(\ref{gammachi}) describe the decay widths 
into light hadrons and into photons or $e^+e^-$ in terms of 46 
NRQCD matrix elements (40 for the $S$-wave decays and $6$ for the $P$-wave
decays). More matrix elements are needed if, as discussed above, 
higher-order operators have to be included.

\subsection{pNRQCD}
\label{secpnrqcd}
The number of non-perturbative parameters may be reduced by integrating out from 
NRQCD degrees of freedom with energy lower than $m$, since each degree of freedom 
that is integrated out leads to a new factorization. Eventually, one ends up with pNRQCD, 
where only degrees of freedom of energy $mv^2$ are left dynamical. In the context of pNRQCD,  
the NRQCD four-fermion matrix elements can be written either as convolutions of Coulomb amplitudes 
with non-local correlators (in the dynamical situation $mv^2 \simg \lQ$) 
or products of wave-functions in the origin by non-local correlators 
(in the dynamical situation $mv^2 \ll \lQ$).  The first situation is believed 
to be the relevant one at least for the bottomonium ground state\cite{long,logs,vill}. 
In the limiting case  $mv^2 \gg \lQ$, the correlators reduce to local condensates and 
explicit formulas have been worked out for the electromagnetic 
decay of the $\Upsilon(1S)$ in\cite{TY}.\footnote{
Concerning the perturbative calculation of the  electromagnetic 
decay width of the $\Upsilon(1S)$ a renormalization group improved expression 
can be found in\cite{PRG} and the wave function in the origin at next-to-next-to-leading 
order in\cite{mel}.} The last situation is expected 
to be the relevant one for most of the existing heavy quarkonia (with the possible 
exception of the bottomonium ground state) and has been studied in\cite{pw,sw,sqrt}.
However, a general consensus on the above assignations of heavy quarkonium
states to dynamical regions has not been reached yet. 
As an example, I mention that in\cite{sumino} it is suggested that also some of the 
higher bottomonium states may be Coulombic bound states while in practically 
all potential models\cite{EichtenQuiggpot,nuci} 
the bottomonium ground state is described by means of confining potentials. 

In the situation $mv^2 \ll \lQ$ and under the condition that:
(a) all higher gluonic excitations between the two heavy quarks develop a mass gap of 
order $\lQ$, (b) threshold effects are small, and (c) contributions coming from 
virtual pairs of quark-antiquark with three momentum of order $\sqrt{m\lQ}$ are subleading,  
\footnote{
Conditions (a) and (b) select the simplest version of pNRQCD with only one degree 
of freedom: the heavy quarkonium singlet field. Condition (a) is supported by lattice 
data on the excitation spectrum of the gluon field around a static quark-antiquark 
pair\cite{JKM}. Condition (b) may be problematic for the $\psi(2S)$, whose mass is very close 
to the $D\bar{D}$ production threshold. Condition (c) is more technical and 
affects the matching to NRQCD. I refer to\cite{sqrt} for a discussion of its validity.}
the NRQCD octet matrix elements relevant for Eqs. (\ref{gamma1})-(\ref{gammachi}) 
can be written at leading order in the $v$ and $\lQ/m$ expansion as\cite{pw,sw}:
\bea
&&
\langle O_8(^3S_1)\rangle_{V_Q(nS)}=
\langle O_8(^1S_0)\rangle_{P_Q(nS)}
\nn\\
&&\qquad\qquad\qquad\qquad\quad
=C_A {|R^{(0)}_{n0}({0})|^2 \over 2\pi}
\left(- {2 (C_A/2-C_F) {\mathcal E}^{(2)}_3 \over 3 m^2 }\right),
\\
&&
\langle O_8(^1S_0)\rangle_{V_Q(nS)}=
{\langle O_8(^3S_1)\rangle_{P_Q(nS)} \over 3}
\nn\\
&&\qquad\qquad\qquad\qquad\quad
=C_A {|R^{(0)}_{n0}({0})|^2 \over 2\pi}
\left(-{(C_A/2-C_F) c_F^2{\mathcal B}_1 \over 3 m^2 }\right),
\\
&&
\langle O_8(^3P_J) \rangle_{V_Q(nS)}=
{\langle O_8(^1P_1)\rangle_{P_Q(nS)} \over 3}
\nn\\
&&\qquad\qquad\qquad\qquad\quad
=(2J+1)\,C_A {|R^{(0)}_{n0}({0})|^2 \over 2\pi}
\left(-{(C_A/2-C_F) {\mathcal E}_1 \over 9 }\right),
\\
&&
\langle O_8(^1S_0)\rangle_{\chi_Q(nJS)} 
= {T_F\over 3}
{\vert R^{(0)\,\prime}_{n1}({0})\vert^2 \over \pi m^2} {\mathcal E}_3, 
\label{matoct}
\eea
where $c_F$ stands for the chromomagnetic matching coefficient, which  
is known at next-to-leading order\cite{ABN}.
Therefore, at the considered order, the octet matrix elements factorize into the product of 
the zeroth-order radial part of the heavy quarkonium wave function, $R_{n\ell}^{(0)}$,  
which may be calculated from the real part of the pNRQCD Hamiltonian\cite{M12},  
and some chromoelectric and chromomagnetic correlators:
\bea
{\cal E}_n = 
{1 \over N_c}\int_0^\infty \!\! dt \, t^n \langle g{\bf E}(t)\cdot g{\bf E}(0)\rangle,
\eea
\bea
{\cal B}_n = 
{1 \over N_c}\int_0^\infty  \!\! dt \, t^n \langle g{\bf B}(t)\cdot g{\bf B}(0)\rangle,
\eea
\bea
{\cal E}^{(2)}_3 &=& {1 \over 4 N_c}
\int_0^\infty  \!\! dt_1\int_0^{t_1} \!\!  dt_2\int_0^{t_2}  \!\! dt_3 \,(t_2-t_3)^3 
\bigg\{
\langle \{g{\bf E}(t_1)\cdot, g{\bf E}(t_2)\}\, \{g{\bf E}(t_3)\cdot, g{\bf 
E}(0)\}\rangle_c
\nn
\\
&&  
~~~~~~~~~~~~~~~~~~~~~~~~~
- {4 \over N_c}\langle {\rm Tr}(g{\bf E}(t_1)\cdot g{\bf E}(t_2))\, {\rm
Tr}(g{\bf E}(t_3)\cdot g{\bf E}(0))\rangle_c
\bigg\} ,
\eea
where
\bea
&&
\langle g{\bf E}(t_1)\cdot g{\bf E}(t_2)\; g{\bf E}(t_3)\cdot g{\bf
E}(0)\rangle_c =
\langle g{\bf E}(t_1)\cdot g{\bf E}(t_2)\; g{\bf E}(t_3)\cdot g{\bf
E}(0)\rangle
\nn
\\
&&
~~~~~~~~~~~~~~~~~~~~~~~~~
-{1 \over N_c} \langle g{\bf E}(t_1)\cdot g{\bf E}(t_2)\rangle\langle
g{\bf E}(t_3)\cdot g{\bf E}(0)\rangle. 
\eea
These correlators are universal in the sense that they do not depend 
on the heavy quarkonium state and, hence, may be calculated once for 
ever, either by means of lattice simulations\cite{correlatorlat}
or specific models of the QCD vacuum\cite{correlator} or extracted from 
some set of experimental data\cite{pw}. 

At leading order in the $v$ and $\lQ/m$ expansion the singlet matrix elements 
can be expressed in terms of the wave functions in the origin only\cite{nrqcd}:
\bea
\label{O1S}
&&\langle O_1(^3S_1)\rangle_{V_Q(nS)}=
\langle O_1(^1S_0)\rangle_{P_Q(nS)}=
\langle O_{\rm EM}(^3S_1)\rangle_{V_Q(nS)}
\nn\\
&&
\qquad\qquad\qquad
=\langle O_{\rm EM}(^1S_0)\rangle_{P_Q(nS)}= C_A {|R^{(0)}_{n0}({0})|^2 \over 2\pi}, \\
\label{O1P}
&&
\langle O_1(^{2S+1}P_J ) \rangle_{\chi_Q(nJS)} = 
\langle O_{\rm EM}(^{2S+1}P_J ) \rangle_{\chi_Q(nJS)}  
={3 \over 2}{C_A \over \pi} |R^{(0)\,\prime}_{n1}({0})|^2.
\label{chio1}
\eea
At leading order the matrix elements of the ${\mathcal P}_1$ operators 
involve also the correlator ${\cal E}_1$:
\bea
&&
\langle {\mathcal P}_1(^3S_1)\rangle_{V_Q(nS)}=
\langle {\mathcal P}_1(^1S_0)\rangle_{P_Q(nS)}=\langle {\mathcal P}_{\rm EM}(^3S_1)\rangle_{V_Q(nS)}
\nn\\
&&
\qquad\qquad\qquad
=\langle {\mathcal P}_{\rm EM}(^1S_0)\rangle_{P_Q(nS)}
=C_A {|R^{(0)}_{n0}({0})|^2 \over 2\pi}
\left(m E_{n0}^{(0)} -{\mathcal E}_1 \right),
\label{P13S1}
\eea
where $E_{n0}^{(0)}\simeq M - 2m$ is the leading-order binding energy.
Equation (\ref{P13S1}) reduces to the formula obtained in\cite{GK} if the 
heavy quarkonium state satisfies also the condition $mv \gg \lQ$.

The leading corrections to the above formulas come from 
quark-antiquark pairs of three momentum of order $\sqrt{m\lQ}$.
The existence of this degree of freedom in the heavy quarkonium system 
has been pointed out in\cite{sqrt}, where the 
leading correction to Eq. (\ref{O1S}) has been calculated.  

The pNRQCD factorization formulas reduce the number of non-perturbative parameters 
needed in order to describe heavy quarkonium decay widths\cite{sw}.
In particular, they have been used to calculate bottomonium matrix elements 
from charmonium data. This is useful since, at the moment, bottomonium data are less abundant than 
charmonium ones. In this way $P$-wave bottomonium inclusive decay widths have been 
calculated\cite{pw,pwphen} before the first data by CLEO-III\cite{CLEOchib} 
were made available. One has to stress, however, that the theoretical uncertainties associated 
to $P$-wave heavy quarkonium decays are rather large, due to the large 
corrections either in the perturbative series (as discussed in Sec. \ref{secper}) 
or in the relativistic expansion (as discussed in Sec. \ref{secrel}). 
For the inclusive decay width of $P$-wave heavy quarkonium 
neither the resummation of large perturbative corrections, nor the computation 
of operators appearing at next-to-leading order in the $v$ and $\lQ/m$ expansion 
has been done yet.

\section{Conclusions}
\label{seccon}
In this letter, I have reviewed some general aspects 
of the theory of inclusive heavy quarkonium decays.
The standard framework is provided by NRQCD and more generally by non-relativistic EFTs of QCD. 
These have put the study of heavy quarkonium observables on the solid ground of QCD.
Models and phenomenological approaches have not necessarily become obsolete: 
they may provide estimates of the non-perturbative paramaters that appear 
in the EFTs. In particular, potential models may still be useful to estimate 
the heavy quarkonium wave functions. However, also potentials are parameters 
of the EFT and have a precise expression in terms of the original degrees of 
freedom (gluons and quarks) of QCD. Lattice gauge theories provide 
the most natural and well founded tool to calculate non-perturbative 
quantities. In fact several lattice determinations 
of matrix elements entering in the heavy quarkonium decay width expression 
at the level of NRQCD, as well as of correlators and Wilson loops 
entering at the level of pNRQCD already exist.

Experimentally, heavy quarkonium decay data have been produced in large 
amount in the last years and have improved the accuracy of several of the 
measured widths and branching ratios. They call for comparable 
precise theoretical determinations. The relevance is twofold.
On one hand we may extract from heavy quarkonium data several 
of the non-perturbative parameters that characterize the low-energy 
dynamics of QCD. This is possible, because we have simple and exact expressions 
that factorize the non-perturbative physics. 
As an example, I mention that the correlators entering in the expression of the decay widths in pNRQCD 
give information on the masses of the heavy quarkonium exotic hybrid states, 
describe the behaviour of the QCD static potential at intermediate distances 
and contribute to the heavy quarkonium levels.
On the other hand we may use heavy quarkonium data to extract 
some of the fundamental parameters of the Standard Model (e.g. the heavy-quark masses 
and $\als$). In the case of $\als$, this is not yet possible from heavy quarkonium
decay data with an accuracy comparable with other determinations due to the difficulties 
discussed in this review\footnote{
The value quoted by the PDG\cite{pdg} seems to underestimate some of the 
uncertainties\cite{pich}.}.  This is one of the many challenges in 
the present and future of heavy quarkonium physics. 

\appendix

\section{Imaginary parts of the dimension 6 and dimension 8 four-fermion matching coefficients}
\label{appA}
The imaginary parts of the four-fermion matching coefficients are known to different 
levels of precision. In the following, I will indicate their most updated values.
The symbols stand for: $C_A = N_c =3$, $C_F = (N_c^2-1)/(2 N_c) = 4/3$, 
$B_F = (C_A/2 - C_F)(C_A^2/2 -2)= 5/12$, $T_F =1/2$, 
$\beta_0 = 11\, N_c/3 - 4\, T_F \, n_f/3$,                                          
$e_Q$ the electrical charge of the quark $Q$ ($e_b=-1/3$, $e_c= 2/3$, ...), 
$\al$ the electromagnetic coupling constant, 
$\als$ the strong coupling constant in the $\MS$ scheme 
and $n_f$ the number of active light flavours 
(typically 3 and 4 for charmonium and bottomonium respectively). 
The scale $\mu$ is the factorization scale and the scale $\mu_R$ the 
renormalization scale. In a physical quantity, like the decay width, 
the $\mu$ dependence will be canceled by low-energy matrix elements 
and the $\mu_R$ dependence by higher-order terms in the perturbative 
expansion. The strong coupling constant $\als(\mu_R)$ has to be understood 
as running with $n_f$ flavours:\footnote{
Note that from
$$
\als(\mu_R) = \als(\mu_R^\prime)\left(
1 + {\als \over \pi} {\beta_0 \over 2} \log {\mu_R^\prime \over \mu_R} 
+ {\cal O}(\als^2) \right), 
$$
and $\als^{(n_f)}(m) = \als^{(n_f+1)}(m)$ it follows that 
$$
\als^{(n_f)}(\mu_R) =  \als^{(n_f+1)}(\mu_R)
\left(1- {2\over 3} {\als \over \pi} T_F \log {\mu_R \over m}
 + {\cal O}(\als^2) \right).
$$
}
\bea
&&
\hspace{-5mm}
{\rm Im\,}f_1(^1S_0) 
\;\;\lower-1.2pt\vbox{\hbox{\rlap{\cite{BCGRswave,Petrelli}}\lower3pt\vbox{\hbox{$\;\;
       =$}}}}\quad\;\;
C_F \left( {C_A\over 2} -C_F \right)  \pi \als(\mu_R)^2
\nn 
\\
&& \times 
\left\{ 1 + {\als \over \pi}\left[ 
\left(-5 + {\pi^2\over 4}\right) C_F 
+ \left({199\over 18} - {13\over 24}\pi^2 \right) C_A 
\right.\right. 
\nn
\\
&& \qquad\qquad
\left.\left.
-{16\over 9} n_f T_F
+ \beta_0 \log {\mu_R\over 2m}
\right]\right\},  
\label{f11S0}
\\
\nn
\\
\nn
\\
&&
\hspace{-5mm}
{\rm Im\,}f_1(^3S_1) 
\;\;\lower-1.2pt\vbox{\hbox{\rlap{\cite{nrqcd,ML}}\lower3pt\vbox{\hbox{$\;
       =$}}}}\quad\;\;
{2\over 9} (\pi^2-9) C_F (C_A^2-4) \left( {C_A\over 2} -C_F \right)^2 \als(\mu_R)^3 
\nn 
\\
&& \times 
\left\{ 1 + {\als \over \pi}\left[ -9.46(2)\, C_F + 4.13(17)\, C_A - 1.161(2)\, n_f 
+ {3\over 2} \beta_0 \log {\mu_R\over m} \right]\right\}
\nn 
\\
&& + \pi e_Q^2 \, \left( \sum_{i=i}^{n_f} e_{q_i}^2\right)\, \alpha^2 
\left\{ 1 - {13\over 4} C_F {\als\over\pi}\right\},  
\label{f13S1}
\\
\nn
\\
&&
\hspace{-5mm}
{\rm Im\,}f_1(^1P_1) 
\;\;\lower-1.2pt\vbox{\hbox{\rlap{\cite{Maltoniphd,huangchao}}\lower3pt\vbox{\hbox{$\;\;
       =$}}}}\quad\;\;
{8 B_F C_F \over 3}  \left( {C_A\over 2} -C_F \right) \als^3 
\left(- {7\over 3} + {7\over 48}\pi^2 - \log
  {\mu \over 2m} \right),
\label{f11P1}
\\
\nn
\\
\nn
\\
&&
\hspace{-5mm}
{\rm Im\,}f_1(^3P_0) 
\;\;\lower-1.2pt\vbox{\hbox{\rlap{\cite{Petrelli}}\lower3pt\vbox{\hbox{$
       =$}}}}\quad\;\;
3 C_F  \left( {C_A\over 2} -C_F \right) \pi \als(\mu_R)^2
\nn 
\\
&& \times 
\left\{ 1 + {\als \over \pi}\left[ 
\left(-{7\over 3} + {\pi^2\over 4}\right) C_F 
+ \left({427\over 81} - {\pi^2\over 144} \right) C_A 
\right.\right.
\nn
\\
&& \qquad\qquad 
\left.\left.
+{4\over 27} n_f\left( -{29\over 6}  - \log {\mu \over 2m}\right) 
+ \beta_0 \log {\mu_R\over 2m}
\right]\right\},  
\label{f13P0a}
\\
&&
\hspace{-5mm}
{\rm Im\,}f_1(^3P_0) 
\;\;\lower-1.2pt\vbox{\hbox{\rlap{\cite{BCGRpwave}}\lower3pt\vbox{\hbox{$
       =$}}}}\quad\;\;
3 C_F \left( {C_A\over 2} -C_F \right)  \pi \als(\mu_R)^2
\nn 
\\
&& \times 
\left\{ 1 + {\als \over \pi}\left[ 
\left(-{7\over 3} + {\pi^2\over 4}\right) C_F 
+ \left({454\over 81} - {\pi^2\over 144} \right) C_A 
\right.\right.
\nn
\\
&& \qquad\qquad 
\left.\left.
+{4\over 27} n_f\left( -{29\over 6}  - \log {\mu \over 2m}\right) 
+ \beta_0 \log {\mu_R\over 2m}
\right]\right\},  
\label{f13P0b}
\\
\nn
\\
\nn
\\
&&
\hspace{-5mm}
{\rm Im\,}f_1(^3P_1) 
\;\;\lower-1.2pt\vbox{\hbox{\rlap{\cite{Petrelli}}\lower3pt\vbox{\hbox{$
       =$}}}}\quad\;\;
{C_F\over 2}  \left( {C_A\over 2} -C_F \right) \als^3 
\left[ \left({587\over 27} - {317\over 144} \pi^2 \right) C_A 
\right.
\nn
\\
&& \qquad\qquad 
\left.
+{8\over 9} n_f\left( -{4\over 3}  - \log {\mu \over 2m}\right) \right],  
\label{f13P1}
\\
\nn
\\
\nn
\\
&&
\hspace{-5mm}
{\rm Im\,}f_1(^3P_2) 
\;\;\lower-1.2pt\vbox{\hbox{\rlap{\cite{Petrelli}}\lower3pt\vbox{\hbox{$
       =$}}}}\quad\;\;
{4\over 5} C_F   \left( {C_A\over 2} -C_F \right) \pi \als(\mu_R)^2
\nn 
\\
&& \times 
\left\{ 1 + {\als \over \pi}\left[ 
-4 C_F 
+ \left({2185\over 216} - {337\over 384} \pi^2 + {5\over 3} \log 2 \right) C_A 
\right.\right.
\nn\\
&& \qquad\qquad 
\left.\left.
+{5\over 9} n_f\left( -{29\over 15}  - \log {\mu \over 2m}\right) 
+ \beta_0 \log {\mu_R\over 2m}
\right]\right\},
\label{f13P2a}
\\
&&
\hspace{-5mm}
{\rm Im\,}f_1(^3P_2) 
\;\;\lower-1.2pt\vbox{\hbox{\rlap{\cite{BCGRpwave}}\lower3pt\vbox{\hbox{$
       =$}}}}\quad\;\;
{4\over 5} C_F \left( {C_A\over 2} -C_F \right) \pi \als(\mu_R)^2
\nn 
\\
&& \times 
\left\{ 1 + {\als \over \pi}\left[ 
-4 C_F 
+ \left({2239\over 216} - {337\over 384} \pi^2 + {5\over 3} \log 2 \right) C_A 
\right.\right.
\nn\\
&& \qquad\qquad 
\left.\left.
+{5\over 9} n_f\left( -{29\over 15}  - \log {\mu \over 2m}\right) 
+ \beta_0 \log {\mu_R\over 2m}
\right]\right\},
\label{f13P2b}
\\
\nn
\\
\nn
\\
&&
\hspace{-5mm}
{\rm Im\,}g_1(^1S_0) 
\;\;\lower-1.2pt\vbox{\hbox{\rlap{\cite{nrqcd}}\lower3pt\vbox{\hbox{$\!
       =$}}}}\quad\;\;
- {4 C_F \over 3} \left( {C_A\over 2} -C_F \right) \pi \als^2,
\label{g11S0}
\\
\nn
\\
\nn
\\
&&
\hspace{-5mm}
{\rm Im\,}g_1(^3S_1) 
\;\;\lower-1.2pt\vbox{\hbox{\rlap{\cite{GK}}\lower3pt\vbox{\hbox{$
       =$}}}}\quad\;\;
- {19\pi^2 -132 \over 54} 
C_F (C_A^2 -4) \left({C_A\over 2} -C_F \right)^2   \als^3,
\label{g13S1}
\\
\nn
\\
\nn
\\
&&
\hspace{-5mm}
{\rm Im\,}f_8(^1S_0) 
\;\;\lower-1.2pt\vbox{\hbox{\rlap{\cite{Petrelli,huangchao}}\lower3pt\vbox{\hbox{$\;\;
       =$}}}}\quad\;\;
B_F \pi \als(\mu_R)^2  
\nn 
\\
&& \times 
\left\{ 1 + {\als \over \pi}\left[ 
\left(-5 + {\pi^2\over 4}\right) C_F 
+ \left({122\over 9} - {17\over 24}\pi^2 \right) C_A 
\right.\right.
\nn
\\
&&
\qquad\qquad 
\left.\left.
-{16\over 9} n_f T_F
+\beta_0 \log {\mu_R\over 2m}
\right]\right\},  
\label{f81S0}
\\
\nn
\\
\nn
\\
&&
\hspace{-5mm}
{\rm Im\,}f_8(^3S_1) 
\;\;\lower-1.2pt\vbox{\hbox{\rlap{\cite{Maltoniphd}}\lower3pt\vbox{\hbox{$
       =$}}}}\quad\;\;
n_f {\pi \als(\mu_R)^2 \over 6}  
\nn 
\\
&& \times 
\left\{ 1 + {\als \over \pi}\left[ 
-{13\over 4} C_F 
+ \left({133\over 18}  + {2\over 3} \log 2 - {\pi^2 \over 4} \right) C_A 
-{10\over 9} n_f T_F  
\right.\right.
\nn\\
&& \qquad\qquad 
\left.\left.
+ \left( - {73\over 4} + {67 \over 36}
  \pi^2 \right){5\over n_f}  
+\beta_0 \log {\mu_R\over 2m}
\right]\right\},  
\label{f83S1}
\\
\nn
\\
\nn
\\
&&
\hspace{-5mm}
{\rm Im\,}f_8(^1P_1) 
\;\;\lower-1.2pt\vbox{\hbox{\rlap{\cite{Maltoniphd}}\lower3pt\vbox{\hbox{$
       =$}}}}\quad\;\;
{C_A\over 12} \pi \als^2,
\label{f81P1}
\\
\nn
\\
\nn
\\
&&
\hspace{-5mm}
{\rm Im\,}f_8(^3P_0) 
\;\;\lower-1.2pt\vbox{\hbox{\rlap{\cite{Petrelli}}\lower3pt\vbox{\hbox{$
       =$}}}}\quad\;\;
3 B_F \pi \als(\mu_R)^2
\nn 
\\
&& \times 
\left\{ 1 + {\als \over \pi}\left[ 
\left(-{7\over 3} + {\pi^2\over 4}\right) C_F 
+ \left({463\over 81}  + {35\over 27} \log 2 - {17\over 216}\pi^2 \right) C_A 
\right.\right.
\nn\\
&& \qquad\qquad 
\left.\left.
+{4\over 27} n_f\left( -{29\over 6}  - \log {\mu \over 2m}\right) 
+\beta_0 \log {\mu_R\over 2m}
\right]\right\},  
\label{f83P0}
\\
\nn
\\
\nn
\\
&&
\hspace{-5mm}
{\rm Im\,}f_8(^3P_1) 
\;\;\lower-1.2pt\vbox{\hbox{\rlap{\cite{Petrelli}}\lower3pt\vbox{\hbox{$
       =$}}}}\quad\;\;
B_F \als^3 \left[ 
\left({1369\over 108} - {23\over 18} \pi^2 \right) C_A 
+{4\over 9} n_f\left( -{4\over 3}  - \log {\mu \over 2m}\right) \right],  
\label{f83P1}
\\
\nn
\\
\nn
\\
&&
\hspace{-5mm}
{\rm Im\,}f_8(^3P_2) 
\;\;\lower-1.2pt\vbox{\hbox{\rlap{\cite{Petrelli}}\lower3pt\vbox{\hbox{$
       =$}}}}\quad\;\;
{4\over 5} B_F  \pi \als(\mu_R)^2
\nn 
\\
&& \times 
\left\{ 1 + {\als \over \pi}\left[ 
-4 C_F 
+ \left({4955\over 431} - {43\over 72} \pi^2 + {7\over 9} \log 2 \right) C_A 
\right.\right.
\nn\\
&& \qquad\qquad 
\left.\left.
+{5\over 9} n_f\left( -{29\over 15}  - \log {\mu \over 2m}\right) 
+ \beta_0 \log {\mu_R\over 2m}
\right]\right\},
\label{f83P2}
\\
\nn
\\
\nn
\\
&&
\hspace{-5mm}
{\rm Im\,}f_{\gamma\gamma}(^1S_0) 
\;\;\lower-1.2pt\vbox{\hbox{\rlap{\cite{BCGRswave,Petrelli}}\lower3pt\vbox{\hbox{$\;\;
       =$}}}}\quad\;\;
\pi e_Q^4 \al^2
\left\{ 1 + {\als \over \pi}
\left(-5 + {\pi^2\over 4}\right) C_F \right\},  
\label{fgg1S0}
\\
\nn
\\
\nn
\\
&&
\hspace{-5mm}
{\rm Im\,}f_{ee}(^3S_1) 
\;\;\lower-1.2pt\vbox{\hbox{\rlap{\cite{lepdec}}\lower3pt\vbox{\hbox{$
       =$}}}}\quad\;\;
{\pi e_Q^2 \al^2 \over 3} \left\{ 1 - 4 C_F {\als(\mu_R) \over \pi} \right.
\nn 
\\
&& 
 + C_F \left({\als \over \pi}\right)^2 
\left[ \left( -{79\over 18} \pi^2  - {2\over 3} \pi^2 \log {\mu\over m}  +
  2 \pi^2 \log 2 + {39\over 4} - \zeta_3 
\right) C_F
\right.
\nn
\\
&& \qquad\qquad 
+ \left( {89\over 72} \pi^2  - \pi^2 \log {\mu\over m}   - 
  {5\over 3}  \pi^2 \log 2  - {151\over 36} - {13\over 2}\zeta_3 
  -{22\over 3} \log {\mu_R \over m}
\right) C_A
\nn
\\
&&  \qquad\qquad 
+ \left( {11\over 9} 
  + {8\over 3} \log {\mu_R \over m}
\right) T_F n_f
\left.
\left.   
+ \left( - {4\over 9}\pi^2  + {44\over 9} 
\right) T_F
\right]
\right\},
\label{fee3S1}
\\
\nn
\\
\nn
\\
&&
\hspace{-5mm}
{\rm Im\,}f_{\gamma\gamma\gamma}(^3S_1) 
\;\;\lower-1.2pt\vbox{\hbox{\rlap{\cite{nrqcd,ML}}\lower3pt\vbox{\hbox{$\;
       =$}}}}\quad\;\;
4 { \pi^2-9 \over  9} e_Q^6 \al^3 
\left\{ 1 - 9.46(2) C_F {\als \over \pi}\right\},  
\label{fggg3S1}
\\
\nn
\\
\nn
\\
&&
\hspace{-5mm}
{\rm Im\,}f_{\gamma\gamma}(^3P_0) 
\;\;\lower-1.2pt\vbox{\hbox{\rlap{\cite{BCGRswave,Petrelli}}\lower3pt\vbox{\hbox{$\;\;
       =$}}}}\quad\;\;
3 \pi e_Q^4 \al^2
\left\{ 1 + {\als \over \pi}
\left(-{7\over 3} + {\pi^2\over 4}\right) C_F  \right\},
\label{fgg3P0}
\\
\nn
\\
\nn
\\
&&
\hspace{-5mm}
{\rm Im\,}f_{\gamma\gamma}(^3P_2) 
\;\;\lower-1.2pt\vbox{\hbox{\rlap{\cite{BCGRswave,Petrelli}}\lower3pt\vbox{\hbox{$\;\;
       =$}}}}\quad\;\;
{4\over 5}  \pi e_Q^4 \al^2
\left\{ 1 - 4 C_F {\als \over \pi}\right\},
\label{fgg3P2}
\\
\nn
\\
\nn
\\
&&
\hspace{-5mm}
{\rm Im\,}g_{\gamma\gamma}(^1S_0) 
\;\;\lower-1.2pt\vbox{\hbox{\rlap{\cite{nrqcd}}\lower3pt\vbox{\hbox{$\!
       =$}}}}\quad\;\;
- {4\over 3} \pi e_Q^4 \al^2,
\label{ggg1S0}
\\
\nn
\\
\nn
\\
&&
\hspace{-5mm}
{\rm Im\,}g_{ee}(^3S_1) 
\;\;\lower-1.2pt\vbox{\hbox{\rlap{\cite{nrqcd}}\lower3pt\vbox{\hbox{$\!
       =$}}}}\quad\;\;
- {4\over 9} \pi e_Q^2 \al^2,
\label{gee3S1}
\\
\nn
\\
\nn
\\
&&
\hspace{-5mm}
{\rm Im\,}g_{ee}(^3S_1,^3D_1) 
\;\;\lower-1.2pt\vbox{\hbox{\rlap{\cite{nrqcd}}\lower3pt\vbox{\hbox{$\!
       =$}}}}\quad\;\;
- {\pi \over 3}   e_Q^2 \al^2.
\label{gee3D1}
\eea
The number over the equal sign indicates the reference/references  
where the most updated value of the matching coefficient can be found.
Some comments are in order. The order $\als$ corrections to 
${\rm Im\,}f_1(^3S_1)$ and ${\rm Im\,}f_{\gamma\gamma\gamma}(^3S_1)$, 
given in Eqs. (\ref{f13S1}) and (\ref{fggg3S1}) respectively, are known 
only numerically and, therefore, affected by a numerical error. 
The last line of Eq. (\ref{f13S1}), proportional to $\alpha^2$, 
comes from the annihilation of the quark-antiquark pair into a virtual photon, 
which then decays into light hadrons. For the order $\als$ corrections to 
${\rm Im\,}f_1(^3P_0)$ and ${\rm Im\,}f_1(^3P_2)$ there are at the moment 
two (numerically slightly) different determinations in the literature. 
Since the question of which one, if any, is correct has not been settled 
yet I have reported both determinations 
(Eqs. (\ref{f13P0a})-(\ref{f13P0b}) and Eqs. (\ref{f13P2a})-(\ref{f13P2b})).
The expression of ${\rm Im\,}f_8(^3S_1)$ given in Eq. (\ref{f83S1}) and taken from\cite{Maltoniphd} 
is different from that one reported in\cite{Petrelli} (there $n_f \to n_f/3$). 
According to one of the authors this is the correct one\cite{maltonipri}.
The electromagnetic matching coefficients (\ref{fgg1S0})-(\ref{gee3D1}) refer 
to annihilation processes where the final states consist of 
two photons (Eqs.  (\ref{fgg1S0}), (\ref{fgg3P0}), (\ref{fgg3P2}) and (\ref{ggg1S0})), 
three photons (Eq. (\ref{fggg3S1})) and 2 massless fermions (Eqs. (\ref{fee3S1}), 
(\ref{gee3S1}) and (\ref{gee3D1})). Note that the matching coefficient 
${\rm Im\,}f_{ee}(^3S_1)$, having been calculated at order $\als^2$, 
is the most accurately known. 

The running equations for the imaginary parts of the matching coefficients of the 
four-fermion NRQCD operators of dimension 6 and 8 have
been obtained in\cite{sw} and can be read there in Appendix C.

\end{document}